# Ground-based Characterization of Hayabusa2 Mission Target Asteroid 162173 Ryugu: Constraining Mineralogical Composition in Preparation for Spacecraft Operations

Lucille Le Corre[1], Juan A. Sanchez[1], Vishnu Reddy[2], Driss Takir[3], Edward A. Cloutis[4], Audrey Thirouin[5], Kris J. Becker[6], Jian-Yang Li[1], Seiji Sugita[7], Eri Tatsumi[7].

[1]Planetary Science Institute, 1700 E Fort Lowell Road, Tucson, Arizona 85719, lecorre@psi.edu.
[2]Lunar and Planetary Laboratory, University of Arizona, 1629 E University Blvd, Tucson, Arizona 85721.
[3]SETI Institute, 89 Bernardo Ave, Suite 200, Mountain View, CA 94043.
[4]Department of Geography, University of Winnipeg, 515 Portage Avenue, Winnipeg, Manitoba, Canada R3B 2E9.
[5]Lowell Observatory, 1400 W Mars Hill Rd, Flagstaff, AZ 86001.
[6]USGS Astrogeology Science Center, 2255 N. Gemini Drive, Flagstaff, AZ 86001.
[7]Dept. of Earth and Planetary Science, School of Science, University of Tokyo, 7-3-1 Hongo, Bunkyo-ku, Tokyo 113-0033, Japan.

**Abstract**

Asteroids that are targets of spacecraft missions are interesting because they present us with an opportunity to validate ground-based spectral observations. One such object is near-Earth asteroid (NEA) (162173) Ryugu, which is the target of the Japanese Space Agency's (JAXA) Hayabusa2 sample return mission. We observed Ryugu using the 3-m NASA Infrared Telescope Facility (IRTF) on Mauna Kea, Hawaii, on July 13, 2016 to constrain the object's surface composition, meteorite analogs, and link to other asteroids in the main belt and NEA populations. We also modeled its photometric properties using archival data. Using the Lommel-Seeliger model we computed the predicted flux for Ryugu at a wide range of viewing geometries as well as albedo quantities such as geometric albedo, phase integral, and spherical Bond albedo. Our computed albedo quantities are consistent with results from Ishiguro et al. (2014). Our spectral analysis has found a near-perfect match between our spectrum of Ryugu and those of NEA (85275) 1994 LY and Mars-crossing asteroid (316720) 1998 BE7, suggesting that their surface regoliths have similar composition. We compared Ryugu's spectrum with that of main belt asteroid (302) Clarissa, the largest asteroid in the Clarissa asteroid family, suggested as a possible source of Ryugu by Campins et al. (2013). We found that the spectrum of Clarissa shows significant differences with our spectrum of Ryugu, but it is similar to the spectrum obtained by Moskovitz et al. (2013). The best possible meteorite analogs for our spectrum of Ryugu are two CM2 carbonaceous chondrites, Mighei and ALH83100.

1.    **Introduction**

Near-Earth asteroid (NEA) (162173) Ryugu is the target of Japanese Space Agency's (JAXA) Hayabusa2 sample return mission. The Apollo-type NEA has a diameter of 865±15 m (Campins et al., 2009; Hasegawa et al., 2008; Müller et al., 2011; Müller et al., 2017) and a geometric albedo of ~0.05 (Ishiguro et al., 2014). Lightcurve observations of the asteroid revealed a synodic rotation period of 7.631±0.005 hours with low amplitude, suggesting a near-spherical shape (Abe et al., 2008). Some visible wavelength spectral observations (Vilas, 2008; Binzel et al., 2001) suggest an absorption feature (~10% deep) centered between 0.6-0.7 µm attributed to $Fe^{2+} > Fe^{3+}$ charge transfer. However, later observations by Moskovitz et al. (2013) and Lazzaro et al. (2013) confirmed that the 0.70 µm feature might not be present or might not be uniform across the surface. The 0.70-µm absorption feature is primarily associated with phyllosilicates (clay minerals) and has been used as a proxy for hydration (Bus and Binzel 2002). The low albedo of Ryugu, coupled with the 0.7-µm absorption feature suggests that Ryugu is rich in organic material and hydrated minerals. Taxonomic classification of the visible spectrum is broadly consistent with a C-type object (Vilas 2008; Binzel et al. 2001). Vilas (2008) noted rotational spectral variations in the 0.7 µm region absorption band; Moskovitz et al. (2013) and Lazzaro et al. (2013) reported no such variations to a precision level of a few percent. They also noted that the asteroid could have localized surface heterogeneities limited to areas less than 5% of the object's surface.

Near-IR spectral observations of Ryugu by Moskovitz et al. (2013) using the SpeX instrument on the NASA IRTF suggests a weak absorption band at 0.9 µm in some of their spectra. Using a curve matching technique, they suggested CM and CI carbonaceous chondrite meteorites as possible analogs for Ryugu. Being very dark, these meteorite types are a good analog for Ryugu because of the very low albedo of the surface. Spectroscopic observations (from 0.4 to 0.85 µm) of Ruygu at different rotation phases presented in Lazzaro et al. (2013) are featureless with little variation. The absorption feature detected at 0.8 µm does not correspond to any expected minerals and is likely an artefact. More recent observations by Perna et al. (2017) in the visible and near-infrared show featureless spectra with a drop-off shortward of 0.45 µm that best resemble spectra of CM and CI carbonaceous chondrites.

The Hayabusa2 spacecraft will rendezvous with Ruygu in mid-2018 and will bring back samples to Earth at the end of 2020 (Watanabe et al., 2017). During its one and a half year-long operations in the vicinity of the asteroid the spacecraft will map the surface with a camera, a spectrometer, a thermal imager and laser altimeter, and deploy three small rovers (MINERVA-II), a lander (MASCOT, Mobile Asteroid Surface Scout), a small carry-on impactor (SCI) and a camera (DCAM3) to monitor the impact experiment. Hayabusa2 will perform sampling at three different sites on the asteroid, one being the material excavated by the SCI impact (Arakawa et al., 2017). The Hayabusa2 project developed this instrument suite to answer fundamental science questions such as: origin and evolution of the in the Solar System, detection of material precursor to life, and understand the origin of water on Earth. More specifically, Hayabusa2 aims at deciphering the evolution of materials in the early Solar System and their subsequent alteration on the asteroid, and understanding the formation of planetesimals (Watanabe et al., 2017).

While Ryugu has been the target of intense ground-based characterization, key questions related to its photometric properties, meteorite analogs and link to other asteroids remain unanswered. This paper presents our modest efforts to address these issues based on spectral observations in 2016 and photometric modeling of archival data.

## 2.    Observations and Data Reduction

We carried out spectral observations of asteroid Ryugu on July 13, 2016 with the 3-m NASA Infrared Telescope Facility (IRTF) on Mauna Kea, Hawaii, when the asteroid was 18.87 visual magnitude, at a phase angle of 13.3°. Near-infrared (NIR) spectra were obtained using the SpeX low-resolution (R~100) spectrometer (Rayner et al., 2003) in prism mode (0.8-2.5 μm). Pairs of images were obtained by alternating the asteroid between two different slit positions (A-B) following the sequence ABBA. Our observational technique involved paring the asteroid with a local solar-type star, which helps modeling the local atmosphere and telluric correction. The placement of these stellar observations, temporally and spatially on the sky in relation to the asteroid, allow us to achieve an optimal correction of the telluric bands. Solar analog star SAO 120107 was also observed to correct for variations in the spectral slope that could result from the use of a non-solar extinction star. Observational circumstances are presented in Table 1. Spectral data was reduced using the IDL-based software Spextools (Cushing et al. 2004). Detailed description of the data reduction procedure can be found in Sanchez et al. (2013) and Sanchez et al. (2015).

**Table 1.** Observational Circumstances for the IRTF Data on July 13, 2016.

| Target | Type | UTC | Airmass | V. Magnitude |
|---|---|---|---|---|
| *(162173) Ryugu* | NEA | 8:50-10:16 | 1.13-1.34 | 18.87 |
| *SAO144137* | Telluric Star | 8:08-10:58 | 1.08-1.53 | 8.88 |
| *SAO120107* | Solar Analog | 5:12-5:24 | 1.04-1.05 | 9.26 |

The rotational period of Ryugu has been estimated at 7.625+/-0.003h (Abe et al. 2008, Kim et al. 2013, Moskovitz et al. 2013), and confirmed by Perna et al. (2017) with an estimate of ~7.63h based on data obtained at the Very Large Telescope (VLT) in 2016. We also saved the MORIS guider images to reconstruct the lightcurve of the asteroid, however due to the faintness of the asteroid the SNR was not adequate enough to reconstruct a lightcurve. In addition, our observations only spanned a fraction of Ryugu's full rotation period. Our first image with MORIS was obtained at 08:18UT, and the last one at 10:52UT. Therefore, our observing block was 2.5h, and we covered less than 1/3 of the full rotation of Ryugu. We phased our data to the Perna et al. (2017) lightcurve and estimated that our spectral observations were obtained around 0.5 phase of their lightcurve.

## 3. Results
### 3.1 Photometric modeling and Reflectance Factor

In this section, we present photometric models of the disk-integrated observations of asteroid Ryugu. The predicted Reflectance Factor at the standard viewing geometry was also derived to be able to directly compare Ryugu's observations to laboratory reflectance measurements of meteorite analogs at the same viewing geometry. This section includes a discussion on the reflectance and albedo quantities relevant for this work, and how viewing conditions [incidence ($i$), emission ($e$), and phase angles ($\alpha$)] vary for each quantity.

Disk-integrated ground-based photometric data of Ryugu are used to constrain the average disk-resolved brightness across Ryugu's surface by fitting phase curve data, compiled by Ishiguro et al. (2014) and references therein. We chose the Lommel-Seeliger model for this work as it has been widely adopted and proven to be able to describe accurately the surfaces of the Moon and the surfaces of dark objects (at least not too close to limb) (Li et al., 2015). The Reflectance Factor (or reflectance coefficient) (*REFF*) is the ratio of the reflectance of the surface to that of a perfectly diffuse (Lambert) surface under the same conditions of illumination (Hapke, 2012). Reflectance, $r_{LS}(i,e,\alpha)$, is directly related to $REFF(i,e,\alpha)$ or $[I/\mathcal{F}](i,e,\alpha)$ as described in the following Lommel-Seeliger $REFF$ function:

$$REFF(i,e,\alpha) = \frac{[I/\mathcal{F}](i,e,\alpha)}{\mu_o} = \frac{\pi r_{ls}(i,e,\alpha)}{\mu_o} = \frac{\varpi_o}{4} \frac{f(\alpha)}{\mu_o + \mu} \quad \text{(eq. 1)},$$

where $\mu_o = \cos(i)$, $\mu = \cos(e)$, $A_{LS} = \frac{\varpi_o}{4\pi}$ is the Lommel-Seeliger albedo, $f(\alpha) = e^{\beta\alpha + \gamma\alpha^2 + \delta\alpha^3}$ is the phase function we adopted (Takir et al. 2015), and $\varpi_o$ is the average particle single scattering albedo. $I$ is the radiance and has units of W/m$^2$/nm/sr. $J = \pi\mathcal{F}$ is the collimated (Sun) light (irradiance) and has units of W/m$^2$/nm.

The minimum and maximum Lommel-Seeliger models capture the uncertainties in the recent computed effective diameter, 865±15 m, by Müller et al. (2017). These models also capture low and high phase angle behavior, and the scatter in the moderate phase angle ground-based observations of (Ishiguro et al. 2014) (Fig. 1). Table 2 shows the models for nominal, maximum, and minimum predicted brightness of Ryugu at 550 nm. We are using the NEAR spacecraft data of Mathilde (Clark et al., 1999) as the best available proxy because ground-based observations of Ryugu are lacking data at the high ends of the phase angle range. Figure 2 shows the predicted flux for Ryugu at a wide range of viewing geometries, using the Lommel-Seeliger model.

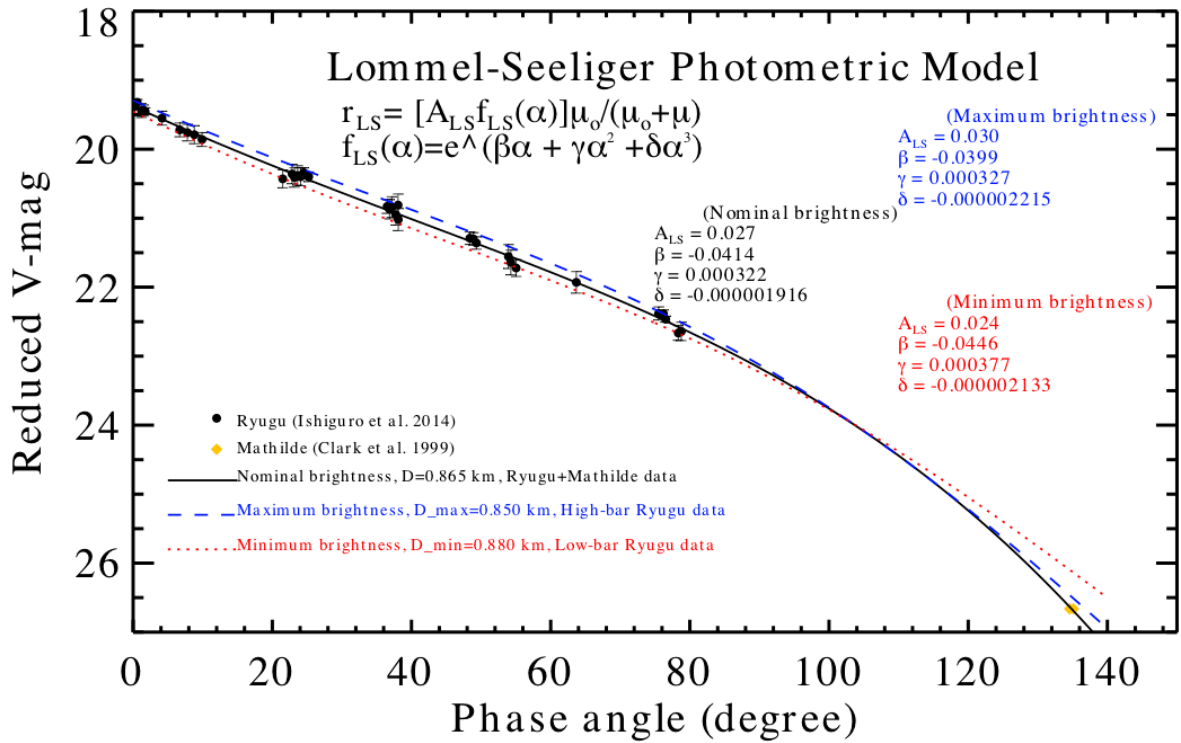

*Figure 1.* *The Reduced V magnitude of Ryugu as a function of phase angle predicted by the Lommel-Seeliger model is shown compared with the ground-based measurements of Ishiguro et al. (2014) and references therein. Shown are the minimum (red dots), maximum (blue dashes), and nominal (black solid line) models. Our nominal model includes the Mathilde data, however, our minimum and maximum models do not because the uncertainties associated with this dataset are not provided in Clark et al. (1999). For the minimum Lommel-Seeliger model we used a diameter value of 880 m; for the maximum Lommel-Seeliger model we used a diameter value of 850 m. Reflectance $r_{LS}$ is in units of $sr^{-1}$.*

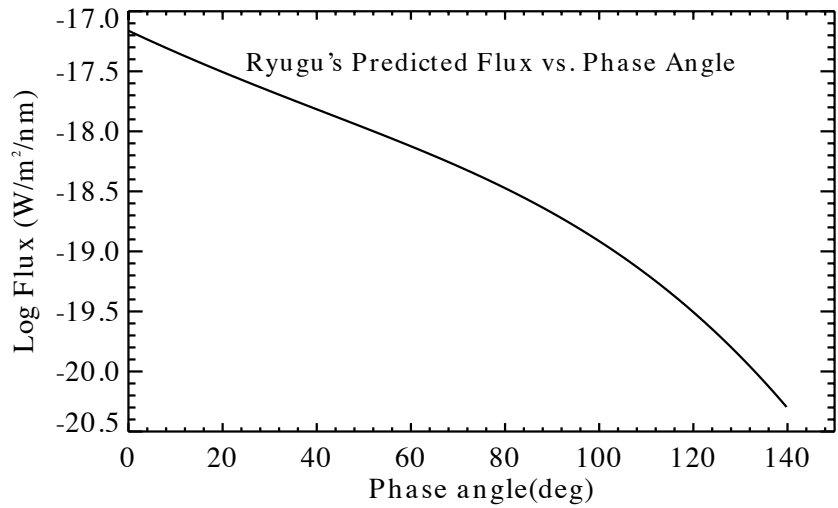

*Figure 2.* *The logarithm of flux ($W/m^2/nm$) derived from the reduced V mag with heliocentric and geocentric distances set at 1 AU, as a function of phase angle, predicted by the Lommel-Seeliger model for Ryugu at 550 nm.*

**Table 2.** Lommel-Seeliger functions that predict [$I/\mathcal{F}$] ($i$, $e$, $\alpha$) (reflectance) of Ryugu at 550 nm. $A_{LS}$ is Lommel-Seeliger albedo and $f(\alpha) = e^{\beta\alpha + \gamma\alpha^2 + \delta\alpha^3}$.

|  | $A_{LS}$ | β | γ | δ |
|---|---|---|---|---|
| *Nominal* | 0.027 | -4.14 x $10^{-2}$ | 3.22 x $10^{-4}$ | 19.16 x $10^{-7}$ |
| *Maximum* | 0.030 | -3.99 x $10^{-2}$ | 3.27 x $10^{-4}$ | 22.15 x $10^{-7}$ |
| *Minimum* | 0.024 | -4.46 x $10^{-2}$ | 3.77 x $10^{-4}$ | 21.33 x $10^{-7}$ |

Table 3 shows Ryugu's *REFF* values at 550 nm, predicted by the Lommel-Seeliger model and computed using Eq. 1, at different laboratory viewing geometries: ($i = 0^o$, $e = 0^o$, $\alpha = 0^o$), ($i = 0^o$, $e = 30^o$, $\alpha = 30^o$), and ($i = 15^o$, $e = 45^o$, $\alpha = 60^o$). These values can be compared directly to laboratory spectra of meteorites sample powders. Laboratory spectra of meteorites are usually obtained at the standard viewing geometry ($i = 0^o$, $e = 30^o$, $\alpha = 30^o$), which corresponds to the value $0.048^{+0.006}_{-0.004}$.

**Table 3.** Ryugu's *REFF* values, predicted by the Lommel-Seeliger model, at different viewing geometries.

| Viewing Geometry | $i = 0^o$, $e = 0^o$, $\alpha = 0^o$ | $i = 0^o$, $e = 30^o$, $\alpha = 30^o$ | $i = 15^o$, $e = 45^o$, $\alpha = 60^o$ |
|---|---|---|---|
| Nominal REFF | 0.042 | 0.048 | 0.040 |
| Maximum REFF | 0.047 | 0.054 | 0.046 |
| Minimum REFF | 0.038 | 0.044 | 0.035 |

Using the Lommel-Seeliger model minimum, maximum, and nominal geometric albedo ($p_v$), phase integral ($q$), and spherical Bond albedo ($A_B$) for Ryugu are computed, following the methodologies described in Takir et al. (2015) (Table 4). The spherical albedo and phase integral are important quantities for computing the bolometric Bond albedo, which will be used for making temperature predictions of the surface of Ryugu for sampling and landing site selection during the Hayabusa2 mission, as well as for thermal inertia calculations to constrain the Yarkovsky effect. The new computed albedo quantities using the Lommel-Seeliger model are consistent with the results of Ishiguro et al. (2014) (Table 4). For a dark object like Ryugu, the normal albedo [*REFF* (0,0,0)] value is close to the geometric albedo because the effects of multiple scattering do not become significant until normal albedo rises above 0.6 (Buratti and Veverka, 1985).

**Table 4.** Albedo quantities of Ryugu computed in this work and by Ishiguro et al. (2014). $p_v$ is the geometric albedo, $q$ is the phase integral, and $A_B$ is the spherical Bond albedo.

| Model | $p_v$ | $q$ | $A_B$ |
|---|---|---|---|
| Lommel-Seeliger (this work) | $0.042^{+0.005}_{-0.004}$ | $0.34^{+0.01}_{-0.01}$ | $0.014^{+0.001}_{-0.001}$ |
| Ishiguro et al. (2014) | $0.047^{+0.003}_{-0.003}$ | $0.32^{+0.03}_{-0.03}$ | $0.014^{+0.002}_{-0.002}$ |

### 3.2 Near-infrared spectrum of Ryugu

The NIR spectrum of Ryugu is shown in Figure 3 (gray circles) along with a smoothed version using the Savitzky-Golay digital filter (red curve), which has the advantage of increasing the S/N without distorting the spectrum. For clarity, in the following discussion the smoothed spectrum will be used for analysis. The spectrum shows no strong well-defined absorption bands over the 0.8-2.5-μm range. There appears to be a slope change near 1.6 μm followed by a broad absorption feature centered near 2.2 μm. There may also be a broad absorption feature centered near 1.0 μm. In comparison to the Moskovitz et al. (2013) spectrum of Ryugu, both show a broad 1-μm region absorption band. The Moskovitz et al. (2013) spectrum is less red than the current spectrum (Figure 4), with 2.5/0.8 μm region reflectance ratios of ~1.3 (Moskovitz et al., 2013) and ~1.6 (this study). The Moskovitz et al. (2013) spectra have lower SNR to confidently identify a broad 2.2 μm region absorption feature, although their spectrum suggests the presence of a narrower absorption band centered at 2.2 μm.

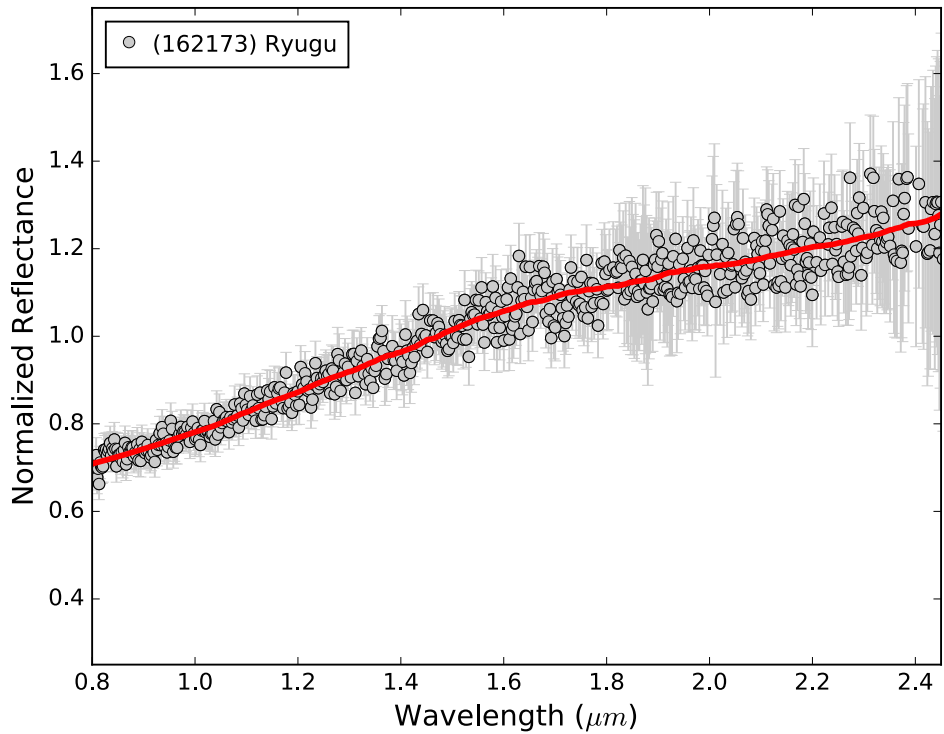

*Figure 3.* *NIR spectrum of asteroid Ryugu at rotation phase 0.5. Also shown, the spectrum of Ryugu after been smoothed (red curve) using the Savitzky-Golay digital filter. The spectrum has been normalized to unity at 1.5 μm.*

Our spectrum of Ryugu seems to differ from the spectra obtained in previous work (Moskovitz et al., 2013; Pinilla-Alonso et al., 2013; Perna et al., 2017), which exhibit a neutral to slightly red spectral slope. Observational factors that have been identified as possible cause for this type of discrepancy include: high humidity and high airmass (>1.6), as well as the spectral type of the extinction star and solar analog used for data reduction. The relative humidity at the time of observation was rather low (~ 10%) and fluctuated between 10 and 13%, with a seeing of 0.81". The airmass was also low and varied from ~1.13-1.34. The extinction star used for telluric correction was the G2V SAO 144137. In addition, the well-known solar analog SAO 120107 was also observed to correct for spectral slope variations that could be introduced by the use of a non-solar extinction star. After taking the ratio between the spectra of the extinction star and that of the solar analog we found that there was little variation in the spectral slope. A second solar analog, Hyades 64 (SAO 93936), which is considered one of the best solar analogs, was also used, and like in the previous case the ratio between the extinction star and this solar analog showed no variation in spectral slope. Therefore, these observational factors do not seem to be the cause for the difference in spectral slope seen among the spectra.

While our spectrum of Ryugu is different from those presented in previous studies, we noticed that it is not unique in terms of overall shape. As example, in Figure 5 we show the smoothed spectrum of Ryugu along with the spectra of (105) Artemis, (85275) 1994 LY and (316720) 1998 BE7. (105) Artemis is a main belt asteroid located at 2.37 AU from the Sun and it is classified as a Ch-type (DeMeo et al., 2009). The spectrum of this object shows a red slope from ~ 0.7 to 1.6 μm, and similarly to Ryugu, it becomes less steep from 1.6 to 2.5 μm. The most obvious difference between both spectra is the more pronounced slope of Ryugu's spectrum. The spectra of near-Earth asteroid (85275) 1994 LY and Mars-crossing asteroid (316720) 1998 BE7 are also very similar to that of Ryugu, with the spectrum of 316720 having an almost perfect match. NEA 85275 has a semimajor axis of ~ 1.9 AU and it likely originated in the inner part of the main asteroid belt, while 316720 has a semimajor axis of ~ 3 AU suggesting a source region far in the outer part of the main belt. The differences in spectral between these asteroids could be due to a combination of factors, including differences in composition, grain size, space weathering, and phase angle. We have also compared the spectrum of Ryugu with that of (302) Clarissa, which was observed with the IRTF for this study on July 03, 2017. Clarissa is a main belt asteroid located at 2.4 AU; it has a geometric albedo of ~ 5% (Tedesco et al. 2004) and it is the parent body of the Clarissa asteroid family (Nesvorny, 2015). Given its location in the main belt, this family has been suggested as a possible source of Ryugu (Campins et al., 2013). The spectrum of this object shares some similarities with that of Moskovitz et al. (2013), but differs from our spectrum of Ryugu (Figure 6), with a neutral to negative slope between ~ 0.8 and 1.1 μm, a weak feature at ~ 1.2 μm, and an increase in slope between 1.2 and 2.5 μm.

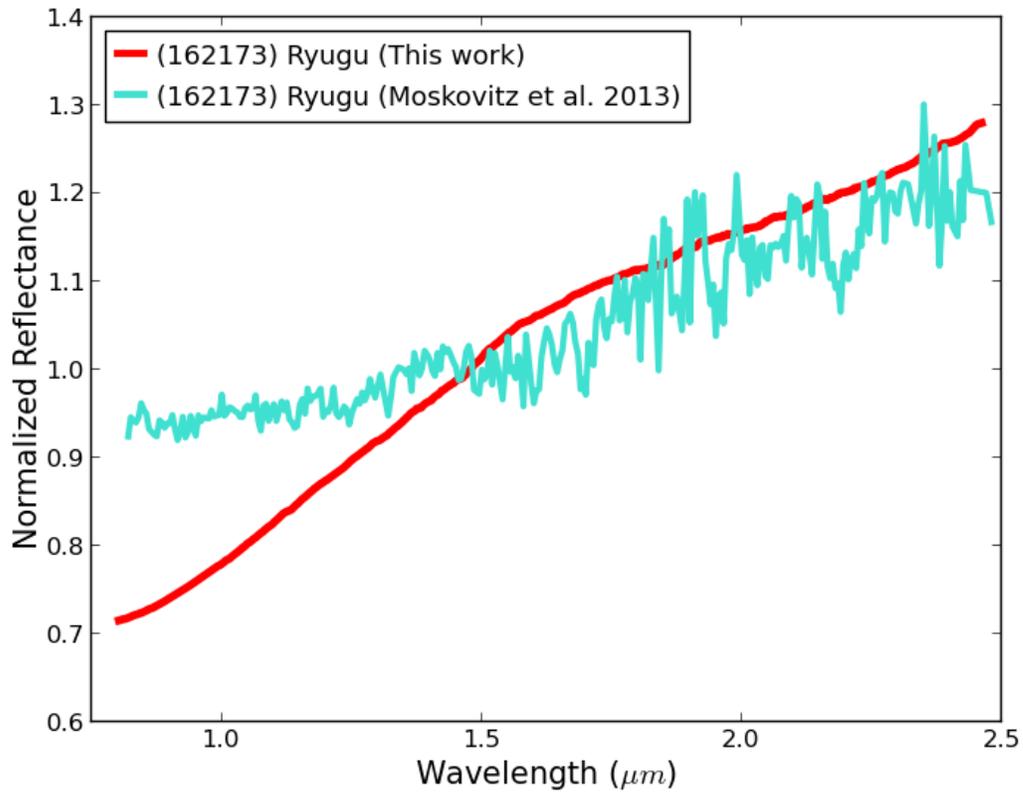

***Figure 4.*** *NIR smoothed spectrum of asteroid Ryugu compared to the spectrum obtained by Moskovitz et al. (2013). Spectra are normalized to unity at 1.5 μm.*

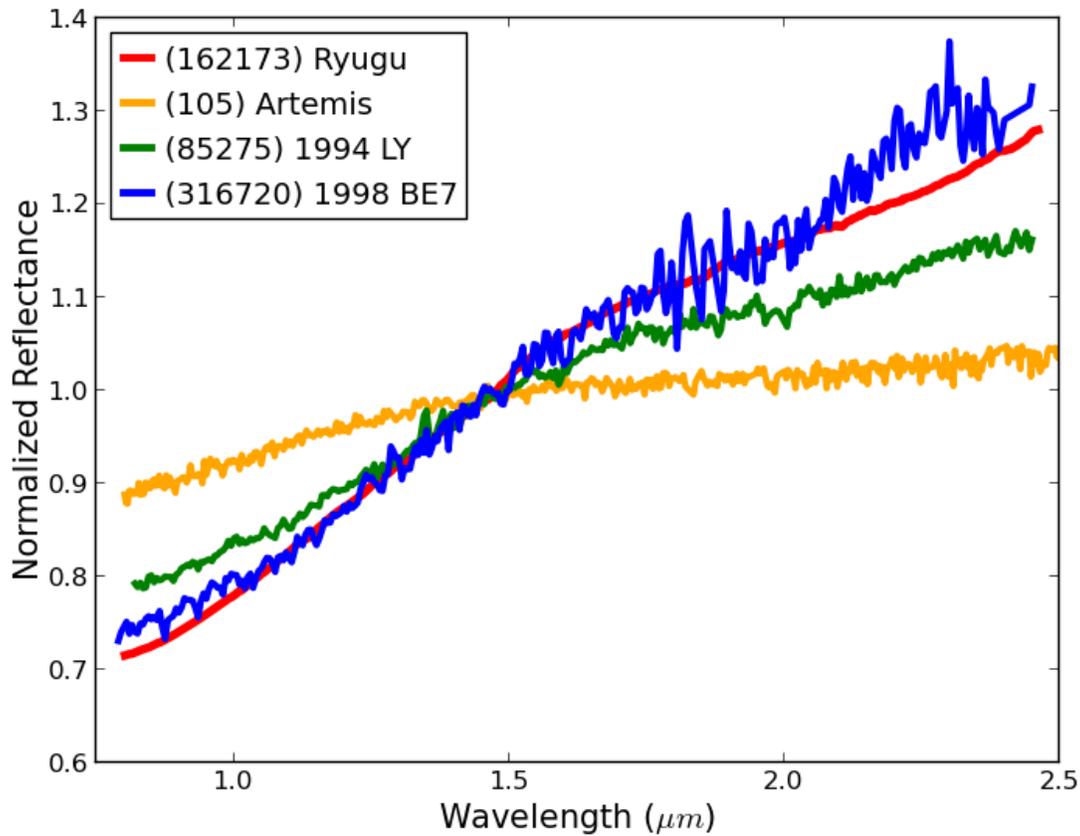

*Figure 5.* *NIR smoothed spectrum of asteroid Ryugu along with the spectra of main belt asteroid (105) Artemis, near-Earth asteroid (85275) 1994 LY and Mars-crossing asteroid (316720) 1998 BE7. The spectrum of Artemis was obtained from Reddy and Sanchez (2016), the spectra of asteroids 85275 and 316720 were obtained from the MIT-Hawaii Near-Earth Object Survey (MITHNEOS) at http://smass.mit.edu/minus.html. All spectra are normalized to unity at 1.5 µm.*

It is not clear why our spectrum of Ryugu shows a steeper spectral slope compared to the spectra obtained by previous studies. Surface heterogeneities could be responsible, although rotationally resolved visible and NIR spectra of Ryugu show no significant variations across the surface of this asteroid (Moskovitz et al., 2013; Perna et al., 2017). We have also considered the possibility that the spectral slope of Ryugu could be influenced by a thermal excess, which is normally seen among dark objects as they get closer to the Sun (Reddy et al., 2012a). Spectra obtained by Pinilla-Alonso et al. (2013) and Perna et al. (2017) show no evidence of such thermal excess which is unusual given the low albedo of Ryugu (~5%) and closer heliocentric distance than our observations.

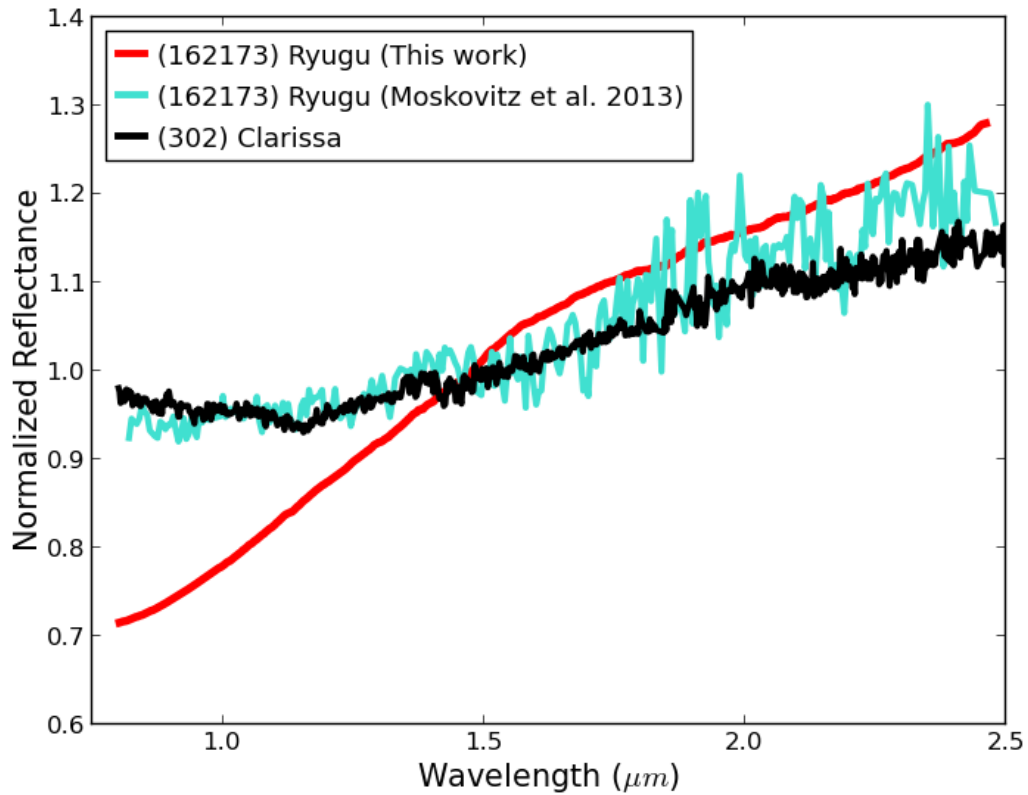

*Figure 6.* NIR smoothed spectrum of Ryugu along with the spectra of this asteroid obtained by Moskovitz et al. (2013) and main belt asteroid (302) Clarissa, observed with the IRTF for this study on July 03, 2017. Spectra are normalized to unity at 1.5 µm.

**3.3 Mineralogy and meteorites analogs**

Even though the 0.8-1.4 µm region spectrum of Ryugu lacks any well-defined absorption bands (Fig. 3), this lack does place constraints on possible meteorite analogs. A number of carbonaceous chondrite groups do show absorption bands in this region. These include: a 0.9 and 1.1 µm pair of absorption bands or an absorption band near 1.0 µm due to Fe-bearing phyllosilicates in some CI (Cloutis et al., 2011a), almost all CM (Cloutis et al., 2011b), and most thermally metamorphosed carbonaceous chondrites (Cloutis et al., 2012a), a 1.05 and 1.25 µm region pair of absorption bands due to olivine in some ureilites (Cloutis et al., 2010), CO (Cloutis et al., 2012b), CV (Cloutis et al., 2012c), and CK chondrites (Cloutis et al., 2012d), and a ~0.9 µm absorption band due to pyroxenes in some ureilites (Cloutis et al., 2010), and CH chondrites (Cloutis et al., 2012e). CO and CV chondrites also exhibit a broad absorption feature in the 2.2 μm region due to $Fe^{2+}$-bearing spinels in calcium aluminum inclusions (CAIs) (Cloutis et al. (2012b, 2012c).

The smoothed spectrum suggests that there may be two broad absorption bands present near 1.0 and 2.2 μm. If both features are present, they are most consistent with CO and CV chondrites (Cloutis et al., 2012b, 2012c). However, reflectance spectra of powdered samples of CO and CV chondrites are generally brighter and less red-sloped than Ryugu. CV chondrites have reflectance from 0.05 to 0.18 at 0.56 μm, and CO chondrites from 0.04 to 0.19, whereas Ryugu has a value of $0.048^{+0.006}_{-0.004}$ at the corresponding viewing geometry. But interestingly, darkening and reddening of carbonaceous chondrites and their associated mafic silicates are possible due to space weathering (Loeffler et al., 2009; Sasaki et al., 2002; Gillis-Davis et al., 2017). Reflectance values at 0.56 μm for CM chondrites (0.03 to 0.12) and for CI chondrites (0.02 to 0.07) are generally the closest to Ryugu's reflectance value among the carbonaceous groups.

We used the online tool M4AST (Popescu et al., 2012) to look for possible meteorite analogs for our spectrum of Ryugu. The best match was obtained for two CM2 carbonaceous chondrites, Mighei (grain size < 40 μm) and ALH83100 (grain size < 100 μm). The spectra of these two meteorites are shown in Figure 7. As can be seen, the shape and slope of the spectra are similar to the spectrum of Ryugu, although the spectrum of ALH83100 seems to have a weak absorption band at ~ 1.15 μm. The reflectance value of these samples at 0.55 μm (0.04 for Mighei and 0.07 for ALH83100) are relatively close to the equivalent reflectance factor calculated for Ryugu ($0.048^{+0.006}_{-0.004}$), with Mighei being the most similar.

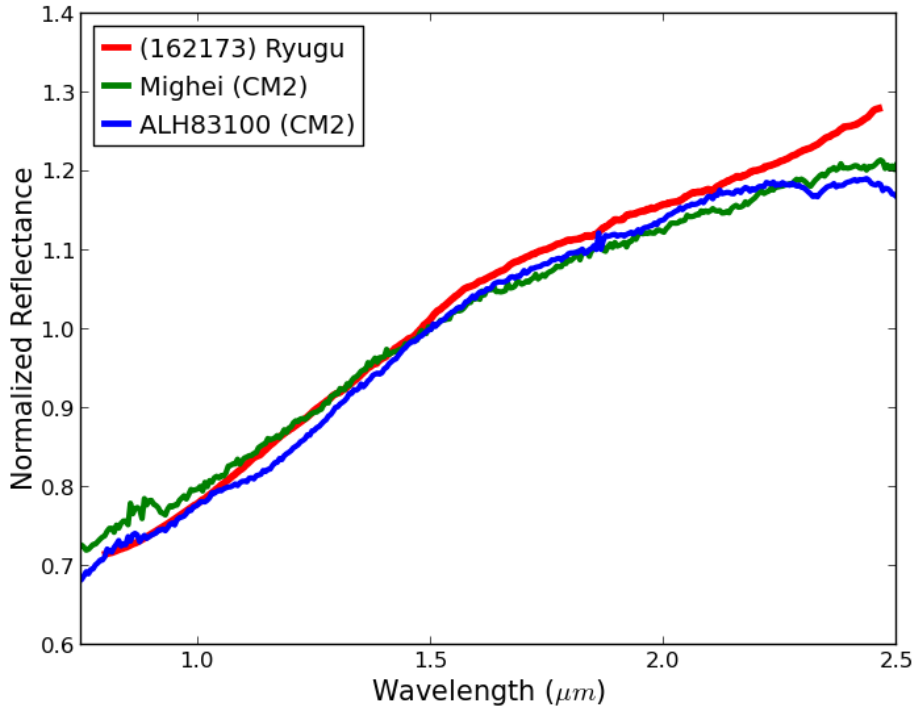

*Figure 7.* NIR smoothed spectrum of asteroid Ryugu and the spectra of CM2 carbonaceous chondrites Mighei (grain size < 40 μm; RELAB ID: MS-CMP-001-D) and ALH83100 (grain size < 100 μm; RELAB IDMB-TXH-051). All spectra are normalized to unity at 1.5 μm.

### 3.4 What to Expect on Ryugu?

While our analysis of ground-based spectral data show that carbonaceous chondrites such as CM2 are good meteorite analogs for Ryugu, surface alteration processes are likely to hinder our ability to confirm this link. Vilas and Gaffey (1989) have identified several dark asteroids in the main belt Cybele and Hilda populations that show a 0.70-μm absorption band similar to Ryugu (Vilas, 2008). Fornasier et al. (2014) also found that this absorption feature has a band depth between 1-7% based on their analysis of 109 dark asteroids. CM2 carbonaceous chondrite meteorites have a similar 0.7-μm band attributed to mixed valence Fe in clay minerals that were formed by aqueous alteration of the asteroid surface. However, the intensity of the 0.7-μm band seems to correlate with the albedo, iron content and temperature. Vilas and Sykes (1996) and Hiroi et al. (2006) noted that the 0.70-μm signature disappeared when carbonaceous chondrite meteorite samples were heated between 400-600°C. In contrast, the 3-μm absorption feature weakens but does not disappear at 600°C. If Ryugu's parent body had undergone thermal alteration (400-600°C) it would be recorded as a variation in the intensity of the 0.7-μm band. Alternatively, the availability of iron in the starting material could also affect the intensity of this band. By correlating the distribution of 0.7-µm and 3.0-µm absorption bands, areas that experienced temperatures <T and >T on Ryugu could be mapped, where T is in the range 400-600°C. Laboratory experiments with CM and CI meteorites by McAdam et al. (2014) showed that not all aqueously altered meteorite exhibit the 0.7-µm absorption band and that the mineralogical changes due to aqueous alteration cannot be derived from the position and depth of this band. Potential meteorite analogs could be identified using color spectra and color maps from ONC-T (Optical Navigation Camera-Telephoto) filters (Kameda et al., 2017), although a more robust link would require Near Infrared Spectrometer (NIRS3) spectral data (Iwata et al., 2017) and sample analysis. Our IRTF spectra of Ryugu will be useful to combine ONC-T multi-band data (0.39-0.95 µm) with NIRS3 spectra (1.8-3.4 µm) that do not have overlapping wavelength ranges.

Surface alteration of material could also occur after the incorporation of material into Ryugu following its parent body break up. The Hayabusa mission to asteroid Itokawa observed two types of surface alterations: possible contamination by exogenic material (dark boulder) (Fujiwara et al., 2006) and space weathering of surface regolith (Hiroi et al., 2006). AMICA (Asteroid Multi Band Camera) observations from the Hayabusa mission also revealed surface albedo variations (~15%) associated with different color units (Sasaki et al., 2007). Exogenic material was previously detected on asteroid surfaces using color filter data on Dawn (e.g., Reddy et al. 2012b, 2012c; Le Corre et al., 2011, 2015). Similarly, we can use the color filter data (albedo and color filter ratios) on ONC-T to identify exogenic meteorite material on the surface of Ryugu. With a geometric albedo of 0.05 (Ishiguro et al., 2014), most silicate-rich meteorites (ordinary chondrites, enstatite achondrites, HEDs) would be easily detectable with ONC-T color filters when emplaced on a dark C-type background material. Moskovitz et al. (2013) noted that the asteroid could have localized surface heterogeneities limited to areas less than 5%. The spatial resolution of ONC-T camera during global mapping (2 m/pixel) is adequate for the identification of small-localized surface variegations due to exogenic impactor material. A 0.7-µm absorption band due to the presence of serpentine was detected in the dark material on the surface of Vesta using the Dawn FC color filters (Nathues et al., 2014). The very low albedo deposits were previously interpreted to be exogenic material brought by a carbonaceous chondrite CM impactor that formed the Veneneia impact basin (Reddy et al., 2012b).

Space weathering refers to processes by which the optical properties of airless body surfaces are altered when exposed to the space environment (Gaffey 2010). On the Moon, space weathering causes an increase in the spectral slope so a reflectance spectrum reddens (increasing reflectance with increasing wavelength). Space weathering also lowers the albedo and decreases the absorption band depths. While lunar-style space weathering is well understood, its extension to asteroids has been less certain. On Vesta, the mixing of exogenic dark material with brighter howarditic Vestan regolith has been termed as space weathering (Pieters et al., 2012). Matsuoka et al. (2015) performed laboratory space weathering experiments on CM2 carbonaceous chondrite meteorites and found that albedo decreases and visible spectral slope became bluer with increased space weathering but at the same time the 0.70 μm and 3.0 μm band depth decreased due to devolatilization of hydrous phases. By correlating the 0.70 μm band depth with visible spectral slope we can constrain "space weathered" regions on Ryugu using ONC-T color filter data. We have used the same methods to identify impact melts on Vesta using Dawn FC color data (Le Corre et al., 2013). Similarly, if a 0.7 μm feature is not detected but a 0.9 μm spectral downturn is present, correlations between albedo, spectral slope and spectral downturn could be used to constrain space weathered regions, as CO/CV chondrites (and olivine) show decreasing albedo, increasing spectral slope, and decreasing mafic silicate band depth, with increasing space weathering.

Kaluna et al. (2016) studied C-complex asteroid families and noted that these asteroids become darker and redder in response to space weathering, as is the case for S-type asteroids. These results do not match the laboratory experiments of Matsuoka et al. (2015) who demonstrated that spectra become bluer with increased space weathering, and the spectral bluing observed by Lantz et al. (2013) and Nesvorny et al. (2005) using ground-based observations. Interestingly, more recent work by Gillis-Davies et al. (2017) on laser-simulated space weathering experiments on CV3 meteorite Allende showed that the weathered meteorite follows the lunar-like spectral changes (spectra redder and darker, absorption band depths decreased as observed by Kaluna et al. (2016)) but also exhibits a different weathering trend for the continuum slope between 450-550 nm (redder, then bluer with increased weathering). According to the experimental work of Lantz et al. (2017) using different types of carbonaceous meteorites, both trends in slope are also possible, with darkening and reddening for some meteorites, and brightening and blueing for others. Both Kaluna et al. (2017) and Lantz et al. (2017) concluded that space weathering trends for carbonaceous asteroids may depend on the original composition of the material. It should be possible to distinguish between the two different trends of space weathering effects on Ryugu with ONC-T color filters by using color ratios and slope calculation for the visible slope in combination with 0.7-μm band depth maps. ONC-T images will allow to validate the proposed space weathering model from Lantz et al. (2017) that predicts no visible space weathering effects in the visible range for a dark asteroid (5-9% albedo) like Ryugu.

## 4.     Summary

We observed (162173) Ryugu, the target of Japanese Space Agency's (JAXA) Hayabusa2 sample return mission, using the 3-m NASA Infrared Telescope Facility (IRTF) on Mauna Kea, Hawaii, on July 13, 2016 when the asteroid was 18.87 visual magnitude, at a phase angle of 13.3°. Our research sheds light on the asteroid's photometric properties, surface composition, meteorite analogs and link to other asteroids in the main belt and NEA populations. Based on our study, we have determined:

- Using the Lommel-Seeliger model we computed the minimum, maximum, and nominal geometric albedo (pv), phase integral (q), and spherical Bond albedo (AB) for Ryugu. The new computed albedo quantities using the model are consistent with the results of Ishiguro et al. (2014).

- We found a near-perfect match between our spectrum of Ryugu and the spectra of near-Earth asteroid (85275) 1994 LY and Mars-crossing asteroid (316720) 1998 BE7, suggesting that their surface regoliths have similar composition.

- We also compared the spectrum of Ryugu with that of main belt asteroid (302) Clarissa, the largest asteroid in the Clarissa asteroid family, suggested as a possible parent body for Ryugu by Campins et al. (2013) based on its low albedo and on dynamical arguments. We found that the spectrum of Clarissa shows significant differences with our spectrum of Ryugu, and it is more similar to the spectrum obtained by Moskovitz et al. (2013).

- The best possible meteorite analogs for our spectrum of Ryugu are two CM2 carbonaceous chondrites, Mighei (grain size < 40 μm) and ALH83100 (grain size < 100 μm).

- We expect the surface regolith of Ryugu to be altered by a range of factors including temperature, contaminated by exogenic material, and space weathering, posing challenges to link spacecraft-ground-based data and sample site selection.

**Acknowledgement**

This work was supported by NASA Hayabusa2 Participating Scientist grant NNX16AK77G (PI: Le Corre). V.R and J.S. work was supported by NASA Near-Earth Object Observations Program grants NNX14AL06G and NNX17AJ19G (PI: Reddy). EAC's work was supported by grants from the Canadian Space Agency, the Natural Sciences and Engineering Research Council of Canada, the Canada Foundation for Innovation, the Manitoba Research Innovation Fund, and the University of Winnipeg. D.T.'s contribution to this work was funded by NASA Hayabusa2 Participating Scientists grants NNX17AL02G. We wish to thank Masateru Ishiguro (Seoul National University) for kindly providing all Ryugu's photometric from Ishiguro et al. (2014). We also thank Nick Moskovitz for sharing the NIR spectrum of Ryugu presented in Moskovitz et al. (2013). All (or part) of the data utilized in this publication were obtained and made available by the MIT-UH-IRTF Joint Campaign for NEO Reconnaissance. The IRTF is operated by the University of Hawaii under contract NNH14CK55B with the National Aeronautics and Space Administration.